\documentclass[12pt]{article}
\usepackage{graphicx}
\usepackage{biblatex}
\usepackage{upgreek}
\newcommand\pubdate{\today}

\def\Title#1{\begin{center} {\Large #1 } \end{center}}
\def\Author#1{\begin{center}{ \sc #1} \end{center}}
\def\Address#1{\begin{center}{ \it #1} \end{center}}

\newcommand\pubblock{\rightline{\begin{tabular}{l}  \\ 
         \pubdate  \end{tabular}}}
\newenvironment{Abstract}{\begin{quotation}  }{\end{quotation}}
\newenvironment{Presented}{\begin{quotation} \begin{center} 
             PRESENTED AT\end{center}\bigskip 
      \begin{center}\begin{large}}{\end{large}\end{center} \end{quotation}}

\begin{document}
\begin{titlepage}
 \pubblock
\vfill
\Title{CMS detector: Run 3 status and plans for Phase-2}
\vfill
\Author{Sre\'cko Morovi\'c}
\Address{on behalf of the CMS Collaboration}

\vfill
\begin{Abstract}
The CMS experiment at the LHC has started data taking in Run 3 at a $\mathrm{pp}$ collision energy of $13.6~\mathrm{TeV}$. In preparation for Run 3, detector systems, such as Pixel Tracker,  HCAL and CSC, have been upgraded due to radiation-induced detector aging or to improve performance. A new GEM muon detector layer was also installed.
The High Luminosity LHC, will provide unprecedented high luminosity and pileup conditions which require more extensive upgrades under the CMS Phase-2 upgrade project. Fully new tracker and high granularity endcap calorimeter will be installed, as well as additional muon detectors and MIP timing layer. Most current detector electronics will be replaced to handle higher triggering rate of up to 750 kHz and increased latency. The upgrades performed for Run 3 as well as planned Phase-2 upgrades and their current development status are summarized.
\end{Abstract}
\vfill
\begin{Presented}
DIS2023: XXX International Workshop on Deep-Inelastic Scattering and
Related Subjects, \\
Michigan State University, USA, 27-31 March 2023 \\
     \includegraphics[width=9cm]{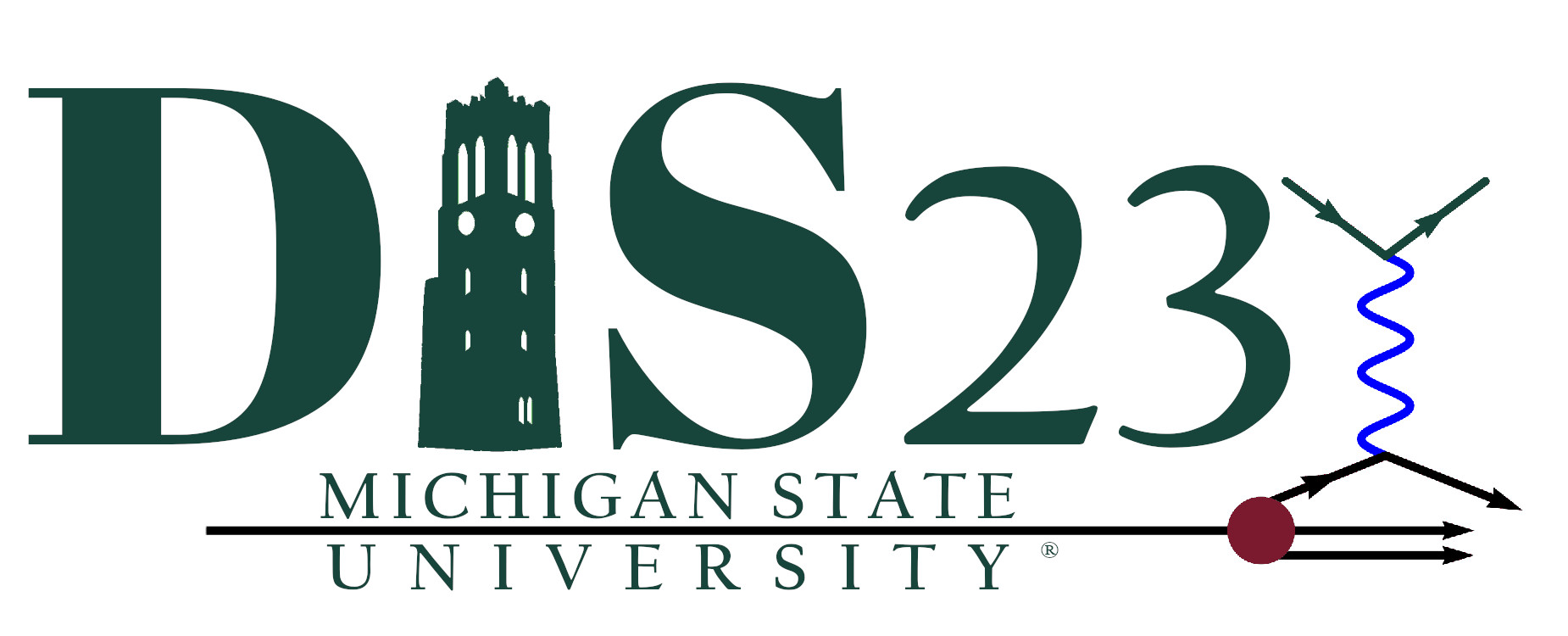}
\end{Presented}
\vfill
\end{titlepage}

\section{Introduction}

The LHC Run 2, during which the CMS detector collected over $140~\mathrm{fb}^{-1}$ proton-proton ($\mathrm{pp}$) collision data at $13~\mathrm{TeV}$, came to a conclusion at the end of 2018 and the Long Shutdown 2 (LS2) period started with the goal of training the magnets for increase in current to allow higher collision energy, as well as installation of scheduled upgrades and maintenance of the large LHC experiments. For CMS~\cite{Chatrchyan:1129810} this involved upgrading several detectors and supporting online systems, which will be described in the following section.
After completion of LS2 in early 2022, Run 3 started with LHC delivering $\mathrm{pp}$ collisions at the new unprecedented center-of-mass energy of $13.6~\mathrm{TeV}$ and instantaneous luminosity of around $2 \times 10^{34}~\mathrm{cm}^2\mathrm{s}^{-1}$. The aim is to collect over $170~\mathrm{fb}^{-1}$ of $\mathrm{pp}$ statistic until 2025 and also include proton-ion and heavy-ion programs.
At the end of Run 3, a three-year Long Shutdown 3 (LS3) period will start, during which the accelerator and experiments will be significantly upgraded and prepared for new running conditions. In 2029, the startup of High Luminosity LHC is planned with the aim of collection of over $3000~\mathrm{fb}^{-1}$ of pp collisions, allowing to probe in more detail the standard model, and perform searches in the context of untested parameter space of new physics models. This will require providing several times larger instantaneous luminosity than in Run 3, approx. $5 - 7 \times 10^{34}~\mathrm{cm}^2\mathrm{s}^{-1}$, and, consequently, harsher conditions for detectors manifested in increased radiation, as well as high rates of proton inelastic scattering (pileup), 140 to 200 interactions per bunch crossing compared to $\approx60$ in Run 3.

\section{Run 3 CMS upgrades}

The Beam Radiation Instrumentation and Luminosity (BRIL) system is responsible for beam and radiation monitoring and protection, as well as luminosity measurement of the experiment. In the view of Run 3, upgrades were mainly aimed at dedicated luminosity measurement systems. 
The pixel luminosity telescope (PLT), consisting of 8 silicon pixel detectors located at each end of CMS, has been refurbished. Fast beam conditions monitor (BCMF1)~\cite{Wanczyk:2827594}, which has been installed, is a redesigned fast particle counter positioned in the proximity of the beam, $6~\mathrm{cm}$ from the beam line and $1.8~\mathrm{m}$ along the beam axis from the interaction point. It provides improved stability, linearity of response and improved radiation tolerance compared to its predecessor.

The inner, pixel tracker of CMS has already been upgraded in Run 2~\cite{thetrackergroupofthecmscollaboration2020cms}, adding the fully new 4-layer layout in 2017. The new tracker had the innermost layer in the barrel region closer to the beam line and overall improved tracking efficiency. During LS2, the innermost layer was again replaced~\cite{Noehte_2022} due to the radiation aging, providing also electronics upgrade which resolved issues with readout synchronization, noise shielding and radiation resistance experienced in Run 2.

The hadronic calorimeter (HCAL) upgrade~\cite{Mans:1481837} involved replacement of photo-sensors and electronics in several HCAL systems. In LS2 it was completed with the barrel region upgrade~\cite{Isik:2810162}. Hybrid photo diodes (HPDs) were replaced with Silicon photomultipliers (SiPM), which have many advantages over HPDs, including high photon detection efficiency, excellent linearity, rapid recovery, better tolerance to radiation and insensitivity to magnetic fields. Radial segmentation was increased, from 2 to 4 in the barrel region, providing improved depth measurement of hadronic showers. The new readout electronics increased readout granularity and redundancy, and improved quality of information sent to the Level-1 (L1) trigger.

In the muon system, consisting of drift tubes (DTs), cathode strip chambers (CSCs), resistive plate chambers (RPCs), and gas electron multipliers (GEMs), a major upgrade was performed on CSC inner rings~\cite{Battilana:2797796}, the CSC layer exposed to the highest flux of particles, where detector electronics were replaced with components compatible with the Phase-2 requirements, supporting readout at the trigger rate up to 750 kHz with latency of 12.5~$\mu\mathrm{s}$. For Run 3 LHC conditions, the occurrence of events lost due to buffer overlow has effectively been eliminated. 

Another important muon upgrade, also part of the staged Phase-2 upgrades, was the installation of the new GEM GE/1 detector ring~\cite{Colaleo:2021453}, one of three planned GEM layers, after a pilot chamber was used in Run 2.  GEM detectors are capable of high-rate, adequate pattern recognition and radiation tolerance, and are particularly useful in background rejection at high particle incidence, which makes these detectors, fitted in the endcap area of the detector, an important upgrade for HL-LHC.

The precise proton spectrometer (PPS), a forward detector consisting of Roman Pots installed close to the beam line,  measures scattered protons in forward regions of the detector. It consists of a tracking detector and timing detectors useful in disentangling multiple collisions (pileup). In Run 3, a new tracker system with 3D silicon pixel  sensors was installed and the timing system received upgraded electronics and diamond detectors~\cite{GarciaFuentes:2811129}. The tracker system provides improved granularity, radiation tolerance and sensors are movable, which allows to distribute the radiation damage.

\subsection{L1 trigger, data acquisition and HLT in Run 3}

The Phase-1 L1 trigger, upgraded~\cite{Tapper:1556311} in 2016, continues to be used in Run-3. 
For Run 3, a demonstration L1-scouting system has been introduced~\cite{Rabady:2816252}. It passively receives L1 trigger data at the full LHC collision rate (nominally 40 MHz, effectively over 30 MHz) via specialized FPGA boards and saves the output for analysis. The advantage of looking at the full-rate trigger information is in allowing searches for rare processes whose signatures are difficult to detect by the trigger.

\begin{figure}[ht]
  \includegraphics[width=0.49\textwidth]{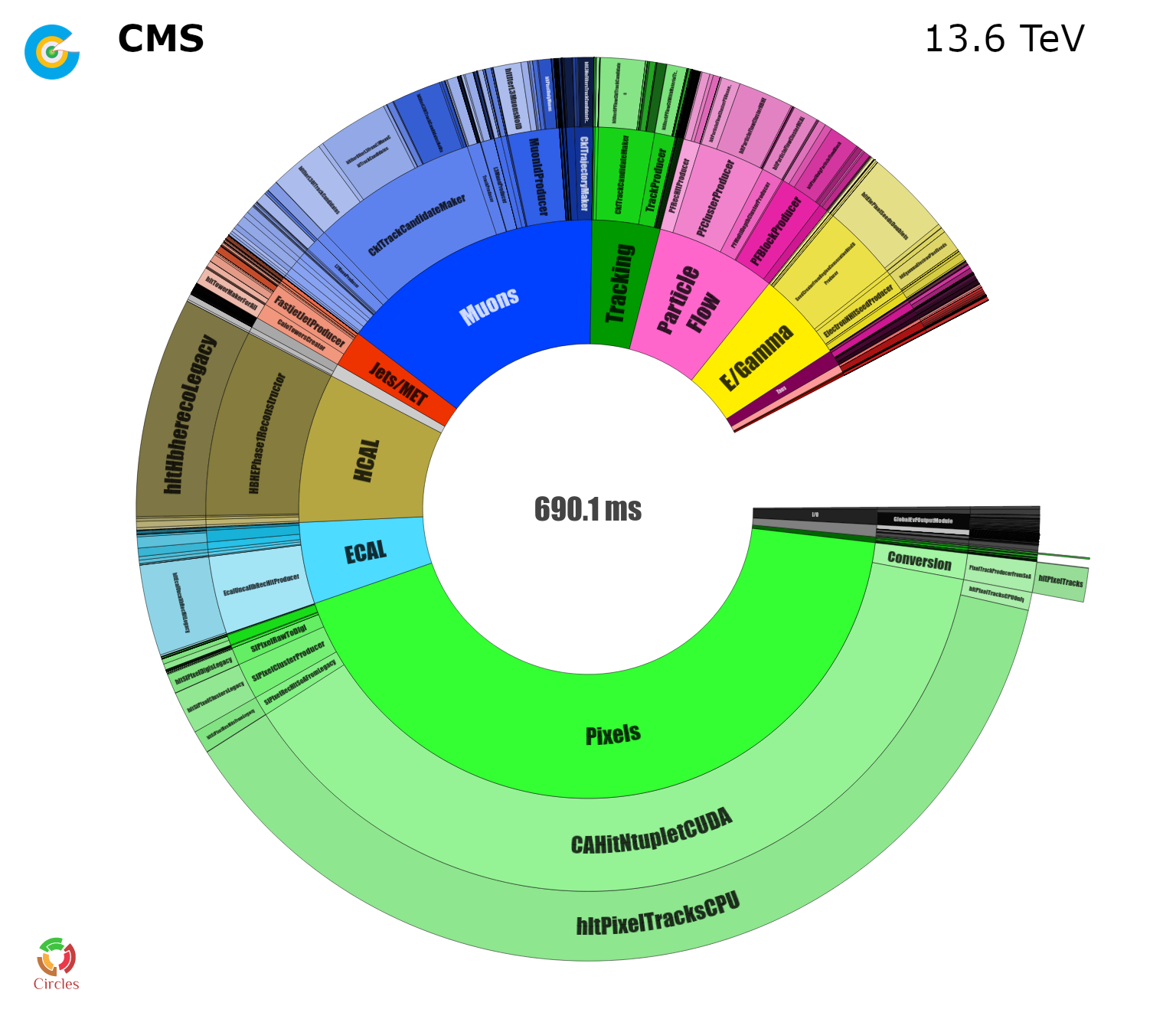}
  \includegraphics[width=0.49\textwidth]{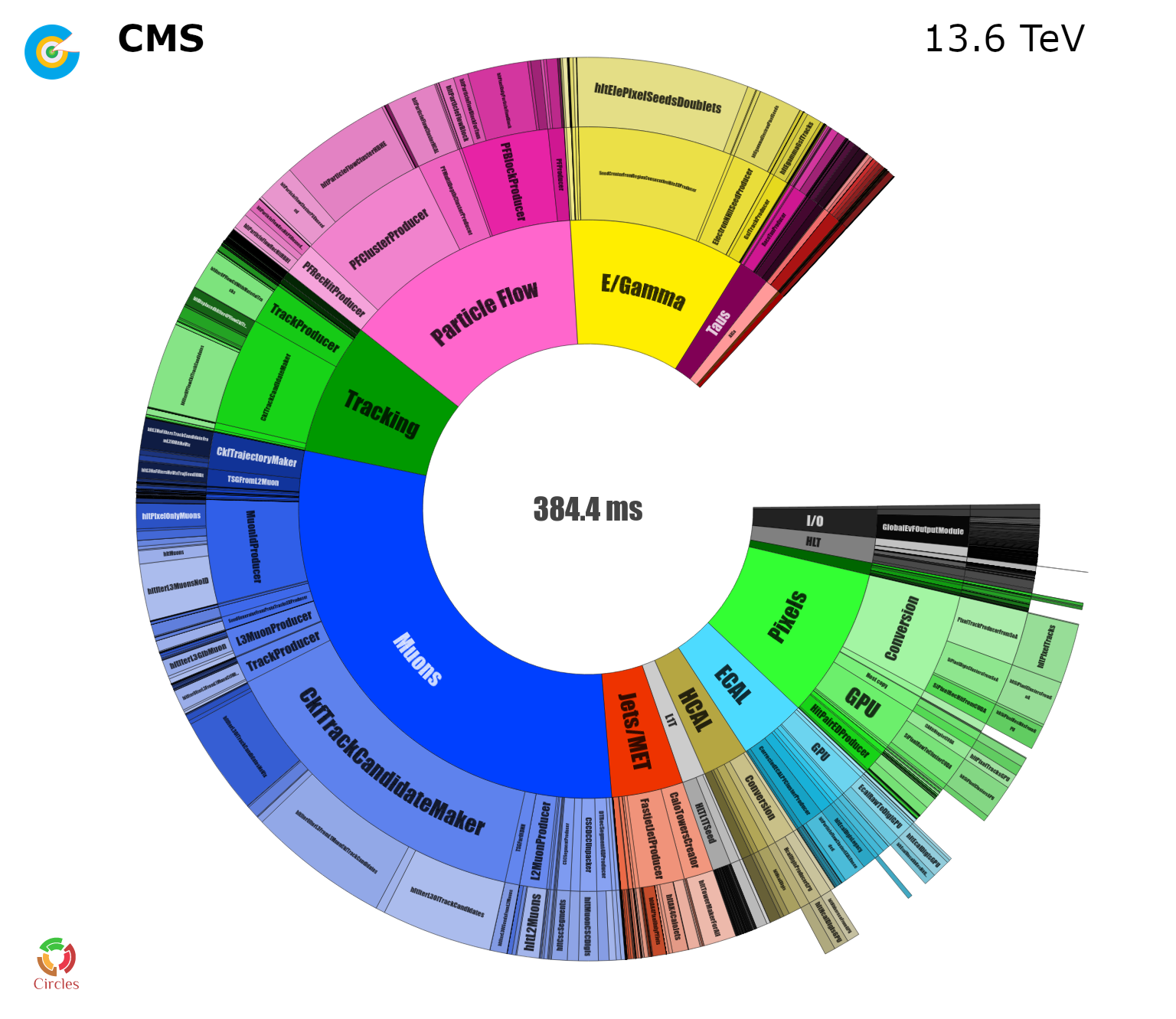}
  \caption{Distribution of the processing time for the HLT reconstruction running only on CPUs (left) and offloading part of the reconstruction to GPUs (right).
  }
  \label{fig:gpus}
\end{figure}

The data acquisition (DAQ) for Run 3 satisfies similar requirements for readout as in Run 2, handling approx. $200~\mathrm{GB/s}$ of data flow at rate above $100 - 110 ~\mathrm{kHz}$ with event sizes of approx. 1.6 MB (at nominal Run 3 LHC conditions). Due to end-of-life of server and network equipment used in Run 2, the system was upgraded with more recent computer and network technologies. Experience with this hardware will also be useful in preparation for Phase-2. A network of Ethernet servers used for data-to-surface transport to the DAQ system has been replaced by a chassis-based $100~\mathrm{Gbit/s}$ Ethernet switch~\cite{Juniper:2021web}, which can flexibly route all TCP traffic from readout cards, FEROLs, read at $10~\mathrm{Gbit/s}$ to  readout servers connected via $100~\mathrm{Gbit/s}$ Ethernet links. Approx. 50 of these nodes, equipped with AMD Rome architecture CPUs, serve also as nodes of the so called folded event builder network architecture~\cite{Mommsen:2649145}, performing the event-building (EVB), a process of collecting all disparate readout event data in one location.
A second $100~\mathrm{Gbit/s}$ chassis-based switch supports the event-building via a RoCE v2 protocol, supporting remote DMA (RDMA) acceleration similar to Infiniband used in Run 2. 
The second chassis switch supports also other DAQ components, such as new Lustre filesystem based storage system as well as data transfer to Tier-0 at CERN for permanent storage.
Fully built events are delivered to the high level trigger (HLT), a cluster of 200 nodes integrated into the DAQ data flow via the same chassis-based switch. 
These nodes are equipped with powerful dual AMD Milan 7763 CPUs, 256 GB RAM, in total amounting to 645 kHS 2006 of processing power.
All HLT nodes are equipped with two Nvidia T4 GPUs and, for the first time, HLT is able to run some of the reconstruction algorithms on GPUs facilitated by the CMS Software framework support for the offloading of computing workloads. Initially this is supported for Pixel, HCAL and ECAL reconstruction with approx. Approximately $40\%$ of the CPU capacity is offloaded to GPUs, as detailed in comparison in Figure~\ref{fig:gpus}.


\section{Operations in Run 3}

After the initial period of commissioning in 2022, 
on July 5th 2022, the run at $13.6~\mathrm{TeV}$ started, lasting until the end of November the same year. In total, around $41 ~\mathrm{fb}^{-1}$ of integrated $\mathrm{pp}$ collision statistics was delivered,
with $37.5~\mathrm{fb}^{-1}$ recorded by CMS ($91.7\%$ efficiency), out of which $34.3~\mathrm{fb}^{-1}$ ($95\%$) was of good data quality and certified for physics analysis.
LHC was performing luminosity leveling at approx. $2 \times 10^{34}~\mathrm{cm}^2 \mathrm{s}^{-1}$, avoiding the high pileup in the beginning of a fill as with the non-leveling scenario, while delivering the integrated luminosity at a comparable rate. In these conditions, the L1 trigger has been accepting events at the rate close to $100~\mathrm{kHz}$ in 2022 and $110~\mathrm{kHz}$ in 2023, and the HLT has been accepting events for permanent storage at above 2 kHz for physics and $2 - 3~\mathrm{kHz}$ for additional data parking.

\section{Phase-2 CMS upgrades}

To retain efficiency and longevity of detectors with increased particle flux and pileup for Phase-2, CMS will undergo a significant upgrade of many detector systems and installation of several new detector designs. Electronics, trigger and readout systems will also undergo significant upgrades.

The inner and outer tracker systems of CMS will undergo a full replacement~\cite{CERN-LHCC-2017-009} with new detectors providing coverage up to pseudorapidity $\eta = 4$. With 20 times more channels than in Run 3, the tracker will provide high-granularity, allowing to retain high resolution and hit efficiency in Phase-2 conditions, and will have high radiation resistance. New RD53A chips~\cite{Missiroli:2839776} have been co-developed with ATLAS for the inner tracker and CMS will use a variant with  $25\times 100~\mathrm{\upmu m}$ sensors. The outer tracker modules are built of two layers of either strips or macro-pixels, providing $100 \mu m$ resolution in $\phi$ and $2.5~\mathrm{cm}$ and $1.5~\mathrm{mm}$, respectively, in $\eta$ dimension.

The MIP timing detector (MTD)~\cite{CMS:2667167} is a planned thin layer of LYSO crystals with SiPM sensors in the barrel, or low gain avalanche diode (LGAD) detectors in endcaps, located between the tracker and calorimeters, which will add precise measurement of production time of MIP particles, with the timing resolution of around $30~\mathrm{ps}$.
This kind of detector, new to CMS, will be useful in disentangling the pileup and will also improve performance of b-jet identification and missing-$p_\mathrm{T}$ measurements. It also enhances sensitivity in searches for long lived particles.

All endcap calorimeters will be replaced with the new high granularity calorimeter (HGCAL)~\cite{CERN-LHCC-2017-023}, a SiPM based detector providing around 6.5 million channels, with the inner Si module tiles having $0.52-1.18~\mathrm{cm}^2$ cells and outer tiles having $4 - 30.25~\mathrm{cm^2}$ cells. The detector will be capable of a 3-dimensional imaging of the particle shower shape, including also precision-timing and energy measurement, and will be integrated with the L1 trigger.
The electromagnetic calorimeter (ECAL) in the barrel region ~\cite{CERN-LHCC-2017-011} will be upgraded with new electronics. PbWO3 crystals in the barrel region are projected to tolerate well high radiation conditions and to retain sufficient transparency, and will continue to be used throughout the Phase-2.

For DT, CSC and RPC muon detectors~\cite{CERN-LHCC-2017-012}, the replacement of on-detector electronics will be performed, satisfying new trigger requirements, and in many cases also improving detector performance. Several new detectors layers will be installed in forward regions. In addition to GEM 1/1 layer already installed, the outer GE 2/1 will be installed. In the inner region, at $\eta = 2.4 - 2.8$, rings of new ME0, 6-layer triple-GEM detectors, will be installed. Improved RPCs (iRPCs), with improved hit rate and time and spatial resolution improvements will be installed in the outer forward region.

BRIL will consist of a total of 15 systems in Phase-2~\cite{BRIL_phase2}. The luminosity precision target is $1\%$ for analysis and $2\%$ for online measurements. A new dedicated luminometer, FBCM, based on Si pad diodes, will provided sub bunch-crossing timing information and high radiation tolerance.

The upgraded L1 Trigger~\cite{CERN-LHCC-2020-004} will accept up to $750~\mathrm{kHz}$ event rate for readout within the $12.5~\mu \mathrm{s}$ latency limit. It will incorporate track-trigger for the outer tracker making use of dual-layer outer tracker modules which will build track stubs. Those objects which will be filtered in an FPGA-based parallelized track-finding system at the full collision rate, allowing separation of low and high $p_\mathrm{T}$ tracks (above 2 GeV). The new trigger will also feature integration of high-granularity calorimetry and will have capability of executing advanced algorithms such as Particle Flow. A production L1-scouting system is planned for Phase-2.

The Phase-2 DAQ system~\cite{Collaboration:2759072} will integrate the new ATCA-based readout system used by detectors, whose electronics will read overall around 50 thousand frontend optical links (lpGBT) from detector into the service cavern, and propagate this data to the the DAQ and Trigger Hub (DTH), comprising dual FPGAs and fast on-board memory. The card will also be responsible for providing precise sub-bunch crossing clock from LHC for synchronization and provide between 4 and 8 input $100~\mathrm{Gbit/s}$ links, as well as the equivalent number of $100~\mathrm{Gbit/s}$ Ethernet links for output to the DAQ data concentrator.
The DAQ system is envisioned to retain the same architecture as in Run 3, but scaled up to handle around 30 times higher bandwidth. It will possibly be based on $400~\mathrm{Gbit/s}$ Ethernet and will take advantage of new generations of computer hardware for I/O nodes. Functionally, the system will perform orbit building instead of collision events which is more optimal for DTH and also reduces the message rate in readout and event-building networks and nodes.
For HLT, new high-granularity detectors as well as high rate of triggered events impose a large, approximately 50-fold increase in computing requirements. Some of the cost of this increase could be mitigated by profiting from evolution of CPUs and GPUs. During Phase-2 the goal is to also migrate 80\% of the workload to co-processors (such as GPUs) which is a promising approach of reducing cost and power consumption of the HLT.  Optimization of the software is another area where improvements are being studied and developed.

\section{Conclusion}
A significant upgrade program was performed in view of Run 3 and CMS has been successfully operating and recording collisions at high efficiency and with good data quality for analysis.
Large R\&D efforts are in progress within the CMS Collaboration in preparation for the Phase-2 upgrade. These efforts are already shifting from engineering and prototyping into the production Phase with the first installations planned for 2026. A large software development effort is also ongoing to integrate new and upgraded detectors into reconstruction and other software used in the collaboration.


\printbibliography

@inproceedings{Isik:2810162,
          author        = "Isik on behalf of the CMS Collaboration, C.",
          collaboration = "CMS",
          title         = "{Phase 1 Upgrade of the CMS Hadron Calorimeter}",
          institution   = "CERN",
          reportNumber  = "CMS-CR-2022-049",
          address       = "Geneva",
          year          = "2022",
          doi           = "10.1016/j.nima.2022.167389",
          %url           = "https://cds.cern.ch/record/2810162",
    }

@techreport{Mans:1481837,
          author        = "CMS Collaboration",
          collaboration = "CMS",
          title         = "{CMS Technical Design Report for the Phase 1 Upgrade of
                           the Hadron Calorimeter}",
          reportNumber  = "CERN-LHCC-2012-015, CMS-TDR-10",
          year          = "2012",
          institution   = "CERN",
          address       = "Geneva",
          url           = "https://cds.cern.ch/record/1481837"
    }

@misc{thetrackergroupofthecmscollaboration2020cms,
          title={The CMS Phase-1 Pixel Detector Upgrade}, 
          author={The Tracker Group of the CMS Collaboration},
          year={2020},
          eprint={2012.14304},
          archivePrefix={arXiv},
          primaryClass={physics.ins-det}
    }

@article{Noehte_2022,
    doi = {10.1088/1748-0221/17/09/C09017},
    year = {2022},
    month = {9},
    publisher = {IOP Publishing},
    volume = {17},
    number = {09},
    pages = {C09017},
    author = {Lars O. S. Noehte and on behalf of the Tracker Group of the CMS collaboration},
    title = {CMS Phase-1 pixel detector refurbishment during LS2 and readiness towards the LHC Run 3},
    journal = {Journal of Instrumentation},
    abstract = {The CMS Phase-1 pixel detector was extracted from the underground cavern after the end of the LHC Run 2 in 2019 and has been kept cold to protect the silicon sensors during the long shutdown period (LS2) in 2019–2021. The LHC is now preparing for the next period of data taking, Run 3, which is scheduled to start in spring 2022. The Phase-1 pixel detector was going through a series of refurbishment and repairs this year to improve the quality of the collected data and enhance the detector performance. The innermost barrel pixel layer has been replaced with new modules and features improved front-end readout chips (PROC600v4), token bit manager chips (TBM10d), and circuit boards to rectify the issues discovered in Run 2. The forward pixel detector has been equipped with new cooling inlets for safe handling and features a revised high-voltage power distribution scheme to better match the low-voltage granularity. All the DC-DC converter modules have been replaced with new modules featuring an improved FEAST2.3 ASIC, which is considerably more robust against the total ionizing dose effect and thus prevents them from breaking during operation. Overall, this article will summarize the refurbishment work of the pixel detector during LS2, and highlight the readiness towards the LHC Run 3 after installation and commissioning.},
    %url = {https://dx.doi.org/10.1088/1748-0221/17/09/C09017}
    }

@techreport{Colaleo:2021453,
          author        = "CMS Collaboration",
          title         = "{CMS Technical Design Report for the Muon Endcap GEM
                           Upgrade}",
          reportNumber  = "CERN-LHCC-2015-012, CMS-TDR-013",
          year          = "2015",
          url           = "https://cds.cern.ch/record/2021453",
    }

@inproceedings{Battilana:2797796,
          author        = "Battilana on behalf of the CMS Collaboration, C.",
          collaboration = "CMS",
          title         = "{Upgrades of the CMS muon detectors: from Run 3 towards HL-LHC. Upgrades of the CMS muon detectors: from Run-3 towards HL-LHC}",
    
          institution   = "CERN",
          reportNumber  = "CMS-CR-2019-243",
          address       = "Geneva",
          year          = "2020",
          doi           = "10.22323/1.364.0156",
          %url           = "https://cds.cern.ch/record/2797796",
    }

@inproceedings{Rabady:2816252,
          author        = "Rabady on behalf of the CMS Collaboration, D. S.",
          collaboration = "CMS",
          title         = "{A 40 MHz Level-1 trigger scouting system for the CMS
                           Phase-2 upgrade}",
          institution   = "CERN",
          reportNumber  = "CMS-CR-2022-102",
          address       = "Geneva",
          year          = "2023",
          doi           = "10.1016/j.nima.2022.167805",
          %url           = "https://cds.cern.ch/record/2816252",
    }

@techreport{Tapper:1556311,
          author        = "CMS Collaboration",
          collaboration = "CMScollaboration",
          editor        = "Tapper, A",
          title         = "{CMS Technical Design Report for the Level-1 Trigger
                           Upgrade}",
          reportNumber  = "CERN-LHCC-2013-011, CMS-TDR-12",
          year          = "2013",
          url           = "https://cds.cern.ch/record/1556311",
    }

@inproceedings{Wanczyk:2827594,
          author        = "Wanczyk on behalf of the CMS Collaboration, J. M.",
          collaboration = "CMS",
          title         = "{Upgraded CMS Fast Beam Condition Monitor for LHC Run 3
                           Online Luminosity and Beam Induced Background
                           Measurements}",
          institution   = "CERN",
          reportNumber  = "CMS-CR-2022-144",
          address       = "Geneva",
          year          = "2022",
          doi           = "10.18429/JACoW-IBIC2022-TH2C2",
          %url           = "https://cds.cern.ch/record/2827594",
    }

@article{GarciaFuentes:2811129,
          author        = "Garcia F. and Francisco I. on behalf of the CMS and LHCf and TOTEM Collaborations",
          title         = "{The TOTEM, CMS PPS and LHCf Run 3 upgrades. 10th Edition
                           of the Large Hadron Collider Physics Conference}",
          year          = "2022",
          url           = "https://cds.cern.ch/record/2811129",
    }

@techreport{CERN-LHCC-2020-004,
          author        = "CMS Collaboration",
          collaboration = "CMS",
          title         = "{The Phase-2 Upgrade of the CMS Level-1 Trigger}",
          institution   = "CERN",
          reportNumber  = "CERN-LHCC-2020-004, CMS-TDR-021",
          address       = "Geneva",
          year          = "2020",
          url           = "https://cds.cern.ch/record/2714892",
    }

@techreport{CERN-LHCC-2017-023,
          author        = "CMS Collaboration",
          collaboration = "CMS",
          title         = "{The Phase-2 Upgrade of the CMS Endcap Calorimeter}",
          institution   = "CERN",
          reportNumber  = "CERN-LHCC-2017-023, CMS-TDR-019",
          address       = "Geneva",
          year          = "2017",
          doi           = "10.17181/CERN.IV8M.1JY2",
          %url           = "https://cds.cern.ch/record/2293646",
    }

@techreport{Collaboration:2759072,
          author        = "CMS Collaboration",
          title         = "{The Phase-2 Upgrade of the CMS Data Acquisition and High
                           Level Trigger}",
          institution   = "CERN",
          reportNumber  = "CERN-LHCC-2021-007, CMS-TDR-022",
          address       = "Geneva",
          year          = "2021",
          url           = "https://cds.cern.ch/record/2759072",
    }

@techreport{CERN-LHCC-2017-011,
          author        = "CMS Collaboration",
          collaboration = "CMS",
          title         = "{The Phase-2 Upgrade of the CMS Barrel Calorimeters}",
          institution   = "CERN",
          reportNumber  = "CERN-LHCC-2017-011, CMS-TDR-015",
          address       = "Geneva",
          year          = "2017",
          url           = "https://cds.cern.ch/record/2283187",
    }

@techreport{CERN-LHCC-2017-012,
          author        = "CMS Collaboration",
          collaboration = "CMS",
          title         = "{The Phase-2 Upgrade of the CMS Muon Detectors}",
          institution   = "CERN",
          reportNumber  = "CERN-LHCC-2017-012, CMS-TDR-016",
          address       = "Geneva",
          year          = "2017",
          url           = "https://cds.cern.ch/record/2283189",
    }

@techreport{CERN-LHCC-2017-009,
          author        = "CMS Collaboration",
          collaboration = "CMS",
          title         = "{The Phase-2 Upgrade of the CMS Tracker}",
          institution   = "CERN",
          reportNumber  = "CERN-LHCC-2017-009, CMS-TDR-014",
          address       = "Geneva",
          year          = "2017",
          doi           = "10.17181/CERN.QZ28.FLHW",
          %url           = "https://cds.cern.ch/record/2272264",
    }

@inproceedings{Missiroli:2839776,
          author        = "Missiroli on behalf of the CMS Collaboration, M.",
          collaboration = "CMS",
          title         = "{Characterisation of the first digital modules with
                           RD53B-CMS readout chips for the Phase-2 Upgrade of the CMS
                           Inner Tracker}",
          institution   = "CERN",
          reportNumber  = "CMS-CR-2022-196",
          year          = "2023",
          doi           = "10.1088/1748-0221/18/01/C01027",
          %url           = "https://cds.cern.ch/record/2839776",
    }

@techreport{CMS:2667167,
          author        = "CMS Collaboration",
          title         = "{A MIP Timing Detector for the CMS Phase-2 Upgrade}",
          institution   = "CERN",
          reportNumber  = "CERN-LHCC-2019-003, CMS-TDR-020",
          address       = "Geneva",
          year          = "2019",
          url           = "https://cds.cern.ch/record/2667167",
    }

@article{Chatrchyan:1129810,
          author        = "CMS Collaboration",
          collaboration = "CMS",
          title         = "{The CMS experiment at the CERN LHC. The Compact Muon
                           Solenoid experiment}",
          journal       = "JINST",
          volume        = "3",
          pages         = "S08004",
          year          = "2008",
          note          = "Also published by CERN Geneva in 2010",
          doi           = "10.1088/1748-0221/3/08/S08004",
          %url           = "https://cds.cern.ch/record/1129810",
    }

@techreport{BRIL_phase2,
          author        = "CMS Collaboration",
          collaboration = "CMS",
          title         = "{The Phase-2 Upgrade of the CMS Beam Radiation,
                           Instrumentation, and Luminosity Detectors: Conceptual
                           Design}",
          institution   = "CERN",
          reportNumber  = "CMS-NOTE-2019-008, CERN-CMS-NOTE-2019-008",
          address       = "Geneva",
          year          = "2020",
          url           = "https://cds.cern.ch/record/2706512",
    }

@techreport{Juniper:2021web,
        author = "{Juniper Networks, Inc.}",
        title = "{QFX10000} modular {Ethernet} switches",
        type = "Datasheet",
        number = "1000529-022-EN",
        year = "2021",
        url = "https://www.juniper.net/us/en/products/switches/qfx-series/qfx10000-modular-ethernet-switches-datasheet.html"
    }

@inproceedings{Mommsen:2649145,
        author = "Mommsen, R. K. and others",
        title = "The {CMS} event-builder system for {LHC} Run 3 (2021--23)",
        booktitle = "{Proc. 23rd International Conference on Computing in High Energy and Nuclear Physics (CHEP 2018): Sofia, Bulgaria, July 9--13, 2018}",
        doi = "10.1051/epjconf/201921401006",
        year = "2019",
        note = "[EPJ Web Conf. 214 (2019) 01006]"
    }

\end{document}